Title:  **How Can Memories Last for Days, Years, or a Lifetime? Proposed Mechanisms for Maintaining Synaptic Potentiation and Memory**

Authors:   Paul Smolen, Douglas A. Baxter, and John H. Byrne

Laboratory of Origin:

Department of Neurobiology and Anatomy

W. M. Keck Center for the Neurobiology of Learning and Memory

McGovern Medical School of the University of Texas Health Science Center at Houston

Houston, Texas 77030

Running Title:   Proposed Mechanisms to Maintain LTP and Memory

Correspondence Address:

Paul D. Smolen

Department of Neurobiology and Anatomy

W.M. Keck Center for the Neurobiology of Learning and Memory

McGovern Medical School of the University of Texas Health Science Center at Houston

Houston, TX 77030

E-mail: Paul.D.Smolen@uth.tmc.edu

Voice: (713) 500-5601

FAX: (713) 500-0623

## Abstract

With memory encoding reliant on persistent changes in the properties of synapses, a key question is how can memories be maintained from days to months or a lifetime given molecular turnover? It is likely that positive feedback loops are necessary to persistently maintain the strength of synapses that participate in encoding. Such feedback may occur within signal-transduction cascades and/or the regulation of translation, and it may occur within specific subcellular compartments or within neuronal networks. Not surprisingly, numerous positive feedback loops have been proposed. Some posited loops operate at the level of biochemical signal transduction cascades, such as persistent activation of $Ca^{2+}$/calmodulin kinase II (CaMKII) or protein kinase M $\zeta$. Another level consists of feedback loops involving transcriptional, epigenetic and translational pathways, and autocrine actions of growth factors such as BDNF. Finally, at the neuronal network level, recurrent reactivation of cell assemblies encoding memories is likely to be essential for late maintenance of memory. These levels are not isolated, but linked by shared components of feedback loops. Here, we review characteristics of some commonly discussed feedback loops proposed to underlie the maintenance of memory and long-term synaptic plasticity, assess evidence for and against their necessity, and suggest experiments that could further delineate the dynamics of these feedback loops. We also discuss crosstalk between proposed loops, and ways in which such interaction can facilitate the rapidity and robustness of memory formation and storage.

**Introduction**

Research over the past 40 years has revealed compelling evidence that memory has different temporal domains. A fundamental difference between short-term memory lasting minutes and long-term memory (LTM) lasting days to weeks or longer is that LTM requires *de novo* protein synthesis and structural changes in neurons (Davis and Squire 1984; Kandel 2001; Mayford et al. 2012). Although changes in protein levels necessary to "support" a memory can certainly persist for many hours if not days, a fundamental enigma is how the physical substrate for storage of LTM can nonetheless be preserved for weeks, months, or a lifetime (Crick 1984; Holliday 1999; Lisman 1985; Roberson and Sweatt 1999; Schwartz 1993; Vanyushin et al. 1974).

At the cellular level, there is now agreement that LTM is stored by a combination of persistent changes in synaptic strength (Lynch 2004) and in intrinsic excitability of neurons (Mozzachiodi and Byrne 2010). In this review we focus on changes in synaptic strength, because mechanisms to maintain synaptic strength have been investigated more thoroughly. Synapses can be strengthened in processes termed long-term potentiation (LTP) at vertebrate synapses and long-term facilitation (LTF) at some invertebrate synapses. Synapses can also be weakened (long-term depression, LTD). These mechanisms contribute to memory formation, and their contribution depends on the type of learning and the brain region involved (Ito et al. 2014; Jörntell 2017; Mirisis et al. 2016; Nabavi et al. 2014; Whitlock 2006). Roles for LTD have been investigated in, for example, cerebellar motor learning (Ito et al. 2014; Jörntell 2017). However, the majority of studies concerned with mechanisms of LTM, and for maintaining LTM for days or longer, have focused on persistence of LTP or LTF. Late LTP (L-LTP) is defined as LTP that lasts beyond ~2 h. L-LTP requires protein synthesis (Abraham and Williams 2008; Frey et al. 1998; Stanton and Sarvey 1984) and requires transcription beyond a time scale of a few hrs (Nguyen et al. 1994). L-LTP in dentate gyrus has been observed to last for at least 1 year (Abraham et al. 2002). LTF persists for at least 1 week, and depends on protein synthesis as well as transcription (Montarolo et al. 1986; Bailey et al. 1992; Miniaci et al. 2008; Hu et al. 2011; Liu et al. 2017).

The majority of experiments delineating molecular mechanisms essential for maintaining LTP, and computational models representing these mechanisms, have focused on excitatory LTP between pyramidal neurons in the hippocampus and in cortical regions (Feldman 2009). Memories first stored in the hippocampus are initially maintained therein by LTP. Subsequently, repeated interactions between hippocampus and other cortical areas reactivate cortical neuronal ensembles that were active during memory encoding, leading to consolidation of memory traces in those cortical areas for long-term storage. This process is referred to as systems consolidation (Frankland and Bontempi 2005; Haist et al. 2001; Kirwan et al. 2008; McGaugh 2000; Nadel and Moscovitch 1997; Squire et al. 2015) and can occur, in part, on a rapid time scale of ~1-2 h (Brodt et al. 2018). We will focus on how LTP and LTM are maintained in the hippocampus for hours or days, and in cortical regions for up to many months. We also review relevant data from other brain regions and from invertebrates such as *Aplysia* and *Drosophila*.

We begin by addressing the likely requirement for synapse-specific positive feedback in maintenance of L-LTP, LTF, and LTM, and the contribution of computational models. We consider proposed synaptic feedback loops, and evidence for or against them. We first discuss feedback that may generate persistent activation of kinases (CaM kinase II, MAP kinase, or protein kinase C isoforms) that increase synaptic strength, and discuss feedback loops that may maintain persistent AMPA receptor phosphorylation, or may persistently increase protein synthesis *via* the translational regulator cytoplasmic polyadenylation element binding protein (CPEB). However, not all feedback loops proposed to play key roles in maintaining LTM are located at synapses. Thus, we proceed to consider putative feedback loops relying on persistent upregulation of gene expression, *via* elevated levels of transcription factors such as cAMP response element binding protein (CREB), or *via* epigenetic regulation. These loops may play an essential permissive role in memory maintenance by enhancing transcription of key genes that produce proteins available to all synapses, and by increasing neuronal excitability. Then we assess evidence for the proposed essential feedback loop that relies on recurrent, ongoing reactivation of assemblies of connected neurons involved in storing a memory. We then discuss crosstalk between proposed feedback loops. The review concludes by discussing how proposed feedback loops operate at multiple spatial levels, from individual kinases to neural assemblies or networks, and by recapitulating evidence supporting the necessity of feedback loops that appear, currently, most likely to be essential for remote (> 1 month) maintenance of LTP and LTM.

**Synapse-specific feedback loops are likely to be essential to maintain long-term potentiation of synapses, preserving LTM**

LTP correlates with structural remodeling at synapses (here referred to as s-LTP), including persistent enlargement and stabilization of postsynaptic dendritic spines. Spine volume and synaptic weight are positively correlated (Cane et al. 2014; El-Boustani et al. 2018). Therefore, LTP induction requires incorporation of new protein into spines. Given that larger spines would be expected to have greater protein turnover, the incorporation of new protein is likely to be persistently increased for long-term maintenance of potentiated spines. Plausibly, some of this new protein is due to enhanced local translation. There is now evidence that local translation is increased at, or close to, dendritic spines after glutamate (Glu) receptor stimulation (Rangaraju et al. 2017). β-actin mRNAs increase localization at or near spines stimulated by Glu uncaging, and a Halo-actin reporter demonstrates concurrent increased levels of Halo-actin at the stimulated region (Yoon et al. 2016). Similarly, after localized dendritic application of a Glu receptor agonist, a Venus-PSD-95 reporter demonstrated enhanced local translation (Ifrim et al. 2015). However, an important caveat is that enhancement of local translation has only been monitored for relatively short times post-stimulus, and has not been demonstrated to last for days or longer at or near potentiated spines. Therefore, although incorporation of new protein at potentiation spines is likely enhanced persistently, a connection to upregulated translation at these time scales has not been established.

This enhanced incorporation, and plausibly enhanced translation, of proteins at potentiated spines during long-term maintenance (days to years) of synaptic strength, is likely to be driven by self-sustaining, synapse-specific positive feedback loops. These loops are likely localized either at potentiated synaptic structures (e.g., enlarged dendritic spines, or immediately adjacent dendritic regions, with persistently activated kinases or upregulated local translation) or at a network level, recurrently reactivating and re-potentiating assemblies of neurons and synapses that store specific memories, as discussed further below. Synapse-specific positive feedback is likely essential because in its absence, given protein turnover time scales of minutes to days (Eden et al. 2011), it is likely that passive relaxation processes, and ongoing homeostatic processes (Abbott and Nelson 2000; Turrigiano et al. 1998), would normalize synaptic strengths to an average value, erasing LTM, LTP, or LTF.

To understand and explore the behavior of putative positive feedback loops and the biochemical signaling pathways that constitute or influence them, it is helpful to use computational models. A common element in models of synaptic positive feedback loops is the presence of a bistable "switch," in which synaptic weight and one or more biochemical variables (*e.g.*, kinase activity) switch from persistent basal values to stimulus-induced, persistently increased values. As an illustration of such a bistable switch, Fig. 1, discussed further below, describes a computational model of one proposed feedback loop in which the activity and synthesis of an atypical protein kinase C isoform, protein kinase M ζ (PKMζ), are persistently elevated following induction of LTP. LTP is represented by an increase in a synaptic weight variable W. Such models are essential to concisely represent feedback loops and the effects of experiments and other perturbations, and to generate empirical predictions that can test the efficacy of proposed feedback mechanisms. We note that it is further possible that interplay of multiple positive feedback loops could generate multistability, with three or more stable values for synaptic weight and biochemical variables (for one such model, see Pi and Lisman 2008). However, current data do not substantially support multistability, which we thus do not consider further.

A number of hypotheses, formalized as computational models, suggest how self-perpetuating kinase activity, or self-perpetuating activation of translation, could sustain positive feedback specifically at strengthened synapses, thus maintaining LTM in hippocampal and cortical areas. In these loops, the regulated output that persistently strengthens synapses is commonly either increased phosphorylation of key synaptic proteins such as AMPA receptors (by persistently active kinases) or increased translation of such proteins.

**Kinase-mediated feedback loops to maintain LTP, LTF, and LTM**

Kinases found to be necessary for the induction of LTP include MAP kinase (MAPK) (English and Sweatt 1997), the MAPK isoform ERK (Rosenblum et al. 2002), CaM kinase II (CaMKII) (Malenka et al. 1989; Otmakhov et al. 1997), the CaM kinase / CaM kinase IV cascade (Ho et al. 2000; Peters et al. 2003), protein kinase A (PKA) (Abel et al. 1997; Matthies and Reymann 1993), classical protein kinase C (PKC) (Colgan et al. 2018), and atypical PKC (Sacktor et al. 1993). Persistent activation of several of these "cognitive kinases" (Schwartz 1993) has been hypothesized as necessary for the long-term maintenance of LTP and LTM. In *Aplysia* and *Drosophila*, homologous kinases are necessary for inducing LTF and LTM: MAPK in *Aplysia* (Martin et al. 1997; Sharma et al. 2003) and *Drosophila* (Pagani et al. 2009), PKA and PKC in *Aplysia* (Lee et al. 2006; Michel et al. 2010), and PKC in the mollusk *Hermissenda* (Farley and Schuman 1991). As discussed below, there is evidence in *Aplysia* that PKA and atypical PKC play roles in maintenance of synaptic strength and LTM.

    1.  Persistent activation of CaM kinase II

One candidate synaptic feedback loop is self-sustaining autophosphorylation of CaMKII. It has been proposed (Lisman and Goldring 1988) that the CaMKII holoenzyme may be well poised to retain self-sustaining activity subsequent to LTP induction. In this positive feedback loop, activated subunits within a dodecameric holoenzyme would phosphorylate neighboring subunits on the Thr286 residue (Thr287 in the β isoform), leading to autonomous activity of those subunits as well. Over time, as existing subunits were degraded, newly incorporated subunits would be phosphorylated by adjacent subunits in the holoenzyme. It was suggested (Lisman and Zhabotinsky 2001) that these persistently active holoenzymes would recruit and stabilize AMPA receptors (AMPARs) at the postsynaptic density (PSD), thus providing a mechanism to maintain synaptic strength and LTP. CaMKII can also upregulate local protein synthesis. Active CaMKII phosphorylates the translation regulator CPEB, upregulating translation of mRNAs containing cytoplasmic polyadenylation elements (CPEs) (Atkins et al. 2004; Atkins et al. 2005).

The CaMKII feedback loop was first formalized in a differential-equation based model demonstrating the plausibility of a bistable switch, in which an LTP-inducing stimulus could persistently change a holoenzyme from a stable inactive form to a self-perpetuating active (autonomous) form (Zhabotinsky et al. 2000). A subsequent model suggested that holoenzyme activity could remain bistable and self-sustaining despite the fluctuations in CaMKII subunit number due to protein degradation and synthesis (Miller et al. 2005). One possible caveat is that autophosphorylation of CaMKII may not always correlate with autonomous activity. In hippocampal slice after tetani or after tetraethylammonium treatment, autophosphorylation of CaMKII was reported to last longer (~ 60 min) than did autonomous CaMKII activity assayed by incorporation of radioactive P into a peptide substrate (Lengyel et al. 2004).

In another hypothesized feedback loop reliant on CaMKII, bistability resulting from feedback between CaMKII autophosphorylation and enhancement of CaMKII translation by CPEB was modeled (Aslam et al. 2009). The mRNA for the α isoform of CaMKII contains CPEs (Wu et al. 1998). Thus, its translation is plausibly upregulated by CPEB, itself activated by CaMKII. The model postulates that the resulting increase in total CaMKIIα and, by mass action, increased CaMKII activity, closes a positive feedback loop that sustains bistability and persistent stimulus-induced elevation of CaMKII activity.

Despite the appeal of these models, empirical support for the persistence of CaMKII activity, as well as experiments designed to block CaMKII activity to test its relationship with maintenance of LTP and memory, have been controversial. It was found (Otmakhov et al. 1997) that inhibition of CaMKII failed to reverse maintenance of hippocampal LTP. However, a different inhibitor did partially reverse maintenance of LTP (Sanhueza et al. 2007). Subsequently, Glu was optically uncaged at single dendritic spines in hippocampal slice, with a stimulus protocol producing persistent spine enlargement, while monitoring CaMKII activity at the stimulated spine and the adjacent dendrite (Lee et al. 2009). Structural LTP (s-LTP), defined as persistent spine enlargement, lasted for > 60 min, but CaMKII activity appeared transient, returning to basal within ~ 2 min in spine and dendrite. These data argue against persistent

CaMKII activation. However, it remains possible that a subpopulation of CaMKII, at the PSD perhaps, below the resolution of these techniques, is persistently active. This issue was addressed (Murakoshi et al. 2017) by expressing a genetically encoded, light-inducible inhibitor of CaMKII in hippocampal slice. Inhibition of CaMKII during stimulus blocked induction of spine s-LTP induced by Glu uncaging, and field LTP induced by tetanus. However, if inhibition of CaMKII was delayed until 1 min after stimulus, normal s-LTP and LTP occurred. These data, together with the observation that synaptic CaMKII is only transiently activated, suggest persistent CaMKII activation is not required to maintain *in vitro* LTP.

However, hippocampal CaMKII phosphorylation persists 20-24 h after inhibitory avoidance (IA) training assayed by immunoblot (Bambah-Mukku et al. 2014; Igaz et al. 2004), suggesting CaMKII may be persistently active. Indeed, in a multi-day IA protocol, mice were trained with two IA sessions on days 1 and 4 (Rossetti et al. 2017). On day 7, a dominant-negative CaMKII construct was expressed *via* injection of a viral vector into the hippocampus. A GFP-alone control was similarly injected. On day 16, LTM was tested. In control mice, memory retention was normal. In mice injected with the construct, memory retention was significantly reduced, suggesting persistent CaMKII activity is necessary to maintain LTM.

In contrast, in *in vivo* experiments (Murakoshi et al. 2017), mice were transfected with the same light-inducible inhibitor used in the LTP experiments discussed above. Inhibition was photoactivated by an optrode for 1 h, either during IA training, or immediately after training. Avoidance memory was assessed after the cessation of inhibition. Memory was blocked by CaMKII inhibition during training, but normal memory was observed following CaMKII inhibition that began immediately after training. These data suggest persistent CaMKII activity is not required to maintain LTM.

How can the inhibition of LTM using a dominant-negative CaMKII construct (Rossetti et al. 2017) be reconciled with the result in which the photoactivatable CaMKII inhibitor does not prevent maintenance of LTP or LTM (Murakoshi et al. 2017)? The possibility that the photoactivatable inhibitor fails to prevent binding of autophosphorylated CaMKII to the NMDA receptor is excluded by data (Murakoshi et al. 2017). However, with this inhibitor (Murakoshi et al. 2017), results were not reported at later times, specifically examining whether CaMKII inhibition immediately post-training affects maintenance of either L-LTP (> 2 h post-stimulus) or of LTM (~24 h post-training). In contrast, using the dominant-negative CaMKII, LTM was maintained at 16 day post training (Rossetti et al. 2017). Thus, it appears possible that a localized pool of persistently active CaMKII, perhaps at the PSD and not resolved by earlier microscopy (Lee et al. 2009), is required specifically for late maintenance of LTP and LTM. Therefore, it would be desirable to extend the CaMKII photoactivatable inhibition experiments to specifically test L-LTP and LTM.

An additional caveat pertains to the observation (Rossetti et al. 2017) that expression of a dominant-negative CaMKII reduced LTM. Overexpressing normal CaMKII suffices to disrupt LTM retrieval (Cao et al. 2008). These authors concluded from additional experiments that retrieval disruption was likely due to active erasure of stored memories by CaMKII overexpression. Given that the molecular mechanism for this erasure is not known, and may not depend on CaMKII activity *per se*, it is possible that expression of dominant-negative CaMKII erases stored LTM. The putative erasure mechanism is not known, but could be related to the observation that overexpression of dominant-negative CaMKII inhibited excitatory synaptic transmission in hippocampal slice (Kabakov and Lisman 2015).

Evidence does suggest that specific interaction of active CaMKII with the NMDAR in the PSD is important for maintaining LTP. LTP induction persistently increases the level of CaMKII / NMDAR complex. Application of a peptide to inhibit formation of the active CaMKII / NMDAR complex blocks LTP induction, and application of this peptide after LTP induction reverses LTP (Sanhueza et al. 2011; Sanhueza and Lisman 2013). Subsequent to LTP reversal, the peptide was removed so that the CaMKII / NMDAR complex could re-form. However, LTP did not fully recover (Sanhueza et al. 2011). This result suggests that LTP maintenance requires, in part, the CaMKII / NMDAR complex. Because this hypothesis requires persistently active CaMKII, in the PSD and possibly below the resolution of earlier microscopy (Lee et al. 2009), further experiments to search for a persistent active CaMKII / NMDAR complex, perhaps using a FRET construct, would be desirable.

Another possibility could be that a structural interaction of CaMKII with another unidentified protein, not dependent on persistent CaMKII phosphorylation, but nonetheless essential for LTP maintenance, was disrupted by competition from dominant-negative CaMKII (Rossetti et al. 2017), but was not disrupted by photoactivatable inhibition (Murakoshi et al. 2017). If so, CaMKII, but not its persistent activity, might still be critical for maintaining LTM.

Overall, the studies reviewed here argue against, but do not completely rule out, a role for persistently self-sustaining CaMKII activity in maintaining LTP and LTM. They do not rule out an essential structural role of CaMKII for maintenance, which might be independent of persistent activity. A recent perspective (Bear et al. 2018) provides additional extensive commentary on the status of the hypothesis that CaMKII is essential for memory storage.

2. Persistent activation of MAPK, PKA, and PKC

Persistent activation of MAPK, PKA, or protein kinase C (PKC), alone or in combination, has been suggested as a mechanism for the preservation of memory. A detailed model was developed (Bhalla and Iyengar 1999) describing dynamics of activation of CaMKII, PKC, MAPK, and other molecules essential for LTP, in hippocampal pyramidal neurons. The model facilitated analysis of possible feedback loops involving these kinases. One simulated positive feedback loop generates bistable PKC and MAPK activity. Stimuli activate PKC by elevating levels of $Ca^{2+}$ and diacylglycerol (DAG). In the model, late PKC activation is sustained due to synergy between elevated levels of arachidonic acid (AA) and basal concentrations of DAG, with AA elevated due to activation of phospholipase A2 by MAPK. The positive feedback loop is closed by activation of the Raf → MAPK signaling cascade by PKC, thus elevating AA, thereby further activating PKC. Experiments in fibroblasts (Bhalla et al. 2002) found bistable MAPK activity, supporting the plausibility of this feedback. Outputs from this feedback loop to maintain increased synaptic strength could include upregulation of local protein synthesis by active MAPK (Kelleher et al. 2004).

However, recent data argue against such persistent PKC activation, and therefore against a necessary role for the above positive feedback loop. Specifically, among the classical PKC isoforms, data indicate that only the α isoform is required for LTP (Colgan et al. 2018). These authors further determined, using Glu uncaging at dendritic spines and a FRET activity reporter, that the activation of the PKC α isoform is transient (< 1 min), not sustained.

Another positive feedback loop has been hypothesized and modeled (Hayer and Bhalla 2005), in which transient activation of PKA, following LTP induction, phosphorylates a critical proportion of AMPA receptors. In this model, kinetic parameters for AMPAR phosphorylation and dephosphorylation are such that following AMPAR phosphorylation, basal PKA activity suffices to persistently maintain elevated AMPAR phosphorylation. Maintained AMPAR phosphorylation, in turn, corresponds to a maintained increase of AMPAR conductance and thus synaptic strength. A recent empirical study optically uncaged Glu at hippocampal dendritic spines (Tang and Yasuda 2017). Glu uncaging produced a transient activation of PKA, at the spine and adjacent dendrite, but activity returned to basal in 5-10 min. These data argue against a hypothesis that persistent PKA activation is required to maintain LTP. However, these data do not disprove the above model in which following transient PKA activation, basal PKA activity suffices to maintain LTP.

In an interesting contrast, LTF induced by the neurotransmitter serotonin (5-HT) in *Aplysia* is accompanied by persistent activation of PKA (Müller and Carew 1998). LTF is attenuated by inhibition of PKA as late as 12 h after LTF induction (Chain et al. 1999; Hegde et al. 1997). These studies provided evidence that persistent PKA activation is mediated by induction of an ubiquitin C-terminal hydrolase, Ap-uch, which facilitates degradation of PKA regulatory subunits, liberating free, active catalytic subunits. The *Ap-uch* promoter contains a variant cAMP response element (CRE) (Mohamed et al. 2005). *Ap-uch* is therefore plausibly induced by the *Aplysia* homologue of mammalian CREB, denoted CREB1a (Bartsch et al. 1998) (herein simplified to CREB1). In turn CREB1 is activated by PKA (Bartsch et al. 1998). This interactions may close a positive feedback loop (active PKA → CREB1 → *Ap-uch* → active PKA) that helps maintain LTF, by sustaining transcription of genes regulated by CREB1 and important for LTF.

For L-LTP, an alternative model for its maintenance posits a bistable positive feedback loop that generates persistent activation of the ERK isoform of MAP kinase (Smolen et al. 2008). However, as noted above, ERK activity was recently monitored in and adjacent to dendritic spines stimulated with local Glu applications (Tang and Yasuda 2017). For ERK, only temporary activation was observed, providing evidence that persistent ERK activity does not occur. These data argue against the operation of the putative positive feedback loops that rely on persistent ERK activation (Bhalla and Iyengar 1999; Smolen et al. 2008). One caveat is that the resolution of these data were not sufficient to resolve differential features, including ERK activation, within a spine. Thus, it remains possible that an ERK (or PKA) sub-population, perhaps at the PSD, could be persistently activated. If a photoactivatable inhibitor of ERK could be delivered to dendrites and spines, this possibility could be excluded by activating the inhibitor ~30 min after stimulus (when ERK activity at a larger scale has returned to baseline) and determining if spine s-LTP was reversed subsequently. This strategy would be similar to that employed for CaMKII (Murakoshi et al. 2017). Interestingly, long-lasting ERK activation (for ~3 hr) does occur during the induction of invertebrate LTF in *Aplysia* (Sharma et al. 2003). ERK activation leads to activation of a CREB kinase, p90 RSK (Philips et al. 2013) and ERK activation is required for LTF (Martin et al. 1997).

3. Persistent activation of protein kinase M ζ (PKMζ)

The hypothesis that a self-sustaining increase in activity and local translation of an atypical PKC isoform, protein kinase M ζ (PKMζ), could underlie long-term maintenance of LTP and various forms of memory, has attracted substantial attention and controversy. PKMζ is persistently active during LTP maintenance (Sacktor et al. 1993), inhibition of PKMζ blocked L-LTP as well as 1 day-old spatial memory (Pastalkova et al. 2006; Serrano et al. 2005), and blocked memory for instrumental and classical conditioning (Serrano et al. 2008). In one study (Shema et al. 2007), the PKMζ inhibitor ZIP was infused 3 d after the induction of LTM, and then LTM was assayed either 1 wk or 1 month later. In both groups LTM was blocked, suggesting that inhibition of PKMζ specifically blocks late maintenance of LTM as opposed to only blocking expression of LTM, and that LTM does not recover after ZIP removal. Similarly, hippocampal LTP was reversed by ZIP infusion and did not recover over 7 following days (Madronal et al. 2010). Translation of the mRNA encoding PKMζ is increased by LTP induction (Hernandez et al. 2003) and PKMζ mRNA localizes to synaptodendritic domains (Muslimov et al. 2004). In mice, hippocampal PKMζ levels were persistently increased for as long as a month after spatial memory formation (Hsieh et al. 2017) and 1 wk after fear conditioning (Wang et al. 2016). Inhibition of an atypical PKC isoform homologous to PKMζ also impaired maintenance of LTM in *Drosophila* (Drier et al. 2002). In *Aplysia*, LTF and long-term sensitization (LTS) of withdrawal reflexes, a form of LTM, are impaired by inhibiting PKM. Impairment occurred when inhibition was applied 3 d after LTF induction (Hu et al. 2017) or as late as 7 d after induction of LTS (Cai et al. 2011), indicating a specific block of late LTM maintenance.

PKMζ can upregulate its own synthesis (Kelly et al. 2007a), forming a positive feedback loop. Models of this positive feedback loop have suggested that bistability, and consequent persistent activation of PKMζ, can be sustained for parameter values consistent with biochemical data (Helfer and Shultz 2018; Smolen et al. 2012). With these models, stochastic simulations were carried out, with fluctuations in PKMζ molecule numbers due to random times of synthesis and degradation. These simulations suggested increased PKMζ activity could persist for weeks or longer given average molecule numbers on the order of hundreds, typical for a large dendritic spine. As an output mechanism for this positive feedback loop, PKMζ promotes trafficking of AMPARs to synapses (Migues et al. 2010), which would increase excitatory postsynaptic current amplitude and thus synaptic strength. PKMζ also promotes translation by repressing a translation inhibitor, Pin1 (Westmark et al. 2010). Figure 1 illustrates a differential equation-based model, and simulations, of a PKMζ feedback loop (Smolen et al. 2012). Figure 1A schematizes the initiation of synaptic PKMζ synthesis by a stimulus modeled as a brief increase in $Ca^{2+}$, with positive feedback from PKMζ then making enhanced PKMζ translation self-sustaining. Figures 1B-C illustrate simulated time courses of variables. PKMζ and the synaptic weight 'W' are bistable due to the positive feedback, thus these variables switch to a persistent upper state post-stimulus. Figure 1D illustrates an ensemble of simulations that extend the model to incorporate stochastic fluctuations in PKMζ molecule numbers within a volume typical of a large dendritic spine. Bistability remains evident and persists despite fluctuations, with the upper state of PKMζ, and therefore W, stable for at least several days.

However, the necessity of the PKMζ positive feedback loop has been challenged. Constitutive PKMζ knockout mice exhibited normal learning and memory in fear conditioning, motor learning, and other protocols (Lee et al. 2013) suggesting PKMζ is not in fact required for LTM. The PKMζ inhibitor ZIP still erased memory in these knockout mice, suggesting ZIP has other targets. Indeed, these authors determined ZIP also inhibited a related PKC isoform, PKCι/λ. These data suggest PKCι/λ is necessary for memory retention in constitutive PKMζ knockout mice. ZIP also reversed LTP in inducible PKMζ knockout mice (Volk et al. 2013). These mice showed normal Schaffer collateral LTP, and did not exhibit deficits in several hippocampal-dependent learning and memory tasks. The importance of demonstrating these results in inducible knockouts is that developmental compensation, in which another kinase such as PKCι/λ may take over an essential role of PKMζ, should be absent in these mice. Thus, these results with the inducible knockout suggest PKMζ is not required for LTP or memory in normal mice.

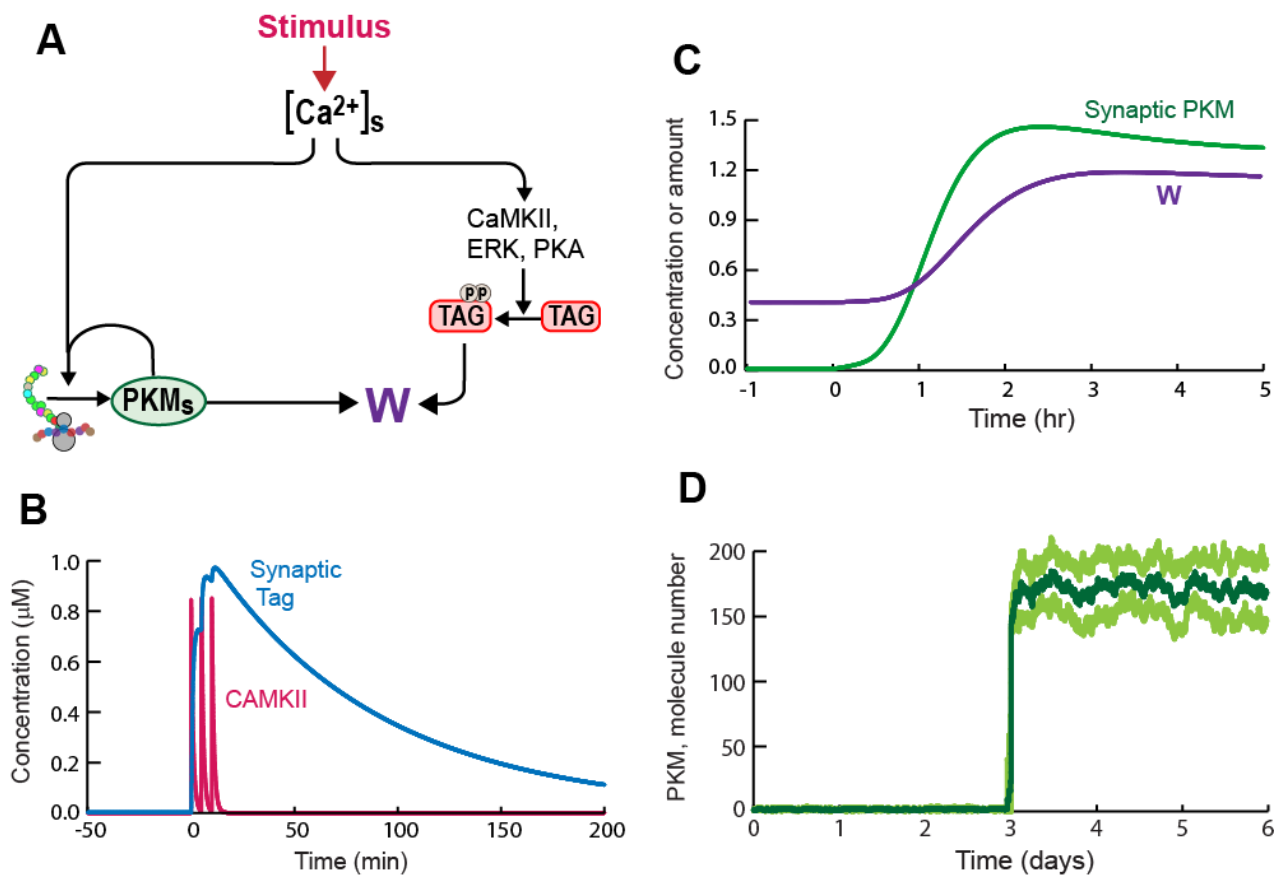

**FIGURE 1  Schematic and simulations of a positive feedback loop based on bistable kinase activity. A,** Stimuli such as a tetanus briefly increase intracellular synaptic $Ca^{2+}$ ($[Ca^{2+}]_S$) and activate synaptic kinases (CaMKII, ERK, PKA). These kinases participate in setting the synaptic tag (Frey and Morris 1997) required for late LTP. Stimuli also enhance translation of synaptic PKMζ, labeled $PKM_S$. In a positive feedback loop, $PKM_S$ further enhances its own translation. In this model (Smolen et al. 2012), brief but strong stimuli switch $PKM_S$ to an upper stable state. $PKM_S$ and the synaptic tag cooperate to enhance synaptic weight W. Thus $PKM_S$ and W are bistable. **B,** simulated time courses, following a tetanus, of model variables which are not bistable, synaptic CaMKII and the synaptic tag. Simulations used the parameter values of Smolen et al. (2012). **C,** time courses of the bistable variables $PKM_S$ and W following a tetanus. Some time courses are vertically scaled for visualization. **D,** an ensemble of stochastic simulations demonstrating robustness of bistability and persistent $PKM_S$ activation to random fluctuations of molecule numbers in a spine volume of 0.2 μm³. Both steady states are stable. Dark green time course, average of $PKM_S$ over 20 simulations. For all simulations the upper state remained stable for at least 3 days. The standard deviation of the 20 trajectories was not large (light green time courses, ±1 SD from average).

These challenges (Lee et al. 2013; Volk et al. 2013) appear to have been addressed in part. Levels of PKCι/λ are persistently increased during LTP maintenance in constitutive PKMζ knockout mice (Tsokas et al. 2016), supporting the importance of PKCι/λ for LTP and learning in these mice. PKMζ antisense RNA blocked late LTP and spatial memory in wild-type mice, but this RNA did not block LTP or memory in the knockout mice (Tsokas et al. 2016). This result indicates that developmental compensation occurred, with PKCι /λ being necessary only in the knockout mice. In addition a PKCι/λ antagonist disrupted late LTP and memory only in the knockout mice. These results suggest that, because developmental compensation is responsible for the necessity of PKCι/λ in the knockout mice, wild-type mice may still require PKMζ for maintenance of LTM, as suggested by the block of late LTP by antisense PKMζ RNA (Tsokas et al. 2016). Thus, these data appear to have addressed the challenge posed by the constitutive knockout data of Lee et al. (2013).

The posited complementary roles of PKCι/λ and PKMζ have been computationally modeled (Jalil et al. 2015). In contrast to the hypothesized persistent upregulation of PKMζ translation from its mRNA, this model posits that maintenance of LTP and LTM by persistently increased PKCι/λ relies on a different type of positive feedback loop, in which PKCι/λ acts to upregulate its own phosphorylation and activity. However, further empirical study is needed to determine whether positive autoregulation of PKCι/λ phosphorylation is sustained, directly or indirectly, during maintenance of LTP. For *Aplysia*, another type of positive feedback loop has been hypothesized, in which PKM activates a calpain that then cleaves the PKC Apl III isoform to generate more PKM (Bougie et al. 2012).

However, the results of Volk et al. (2013) using an inducible knockout, in which developmental compensation by PKCι/λ should not occur, are still in contradiction to the hypothesis that PKMζ is essential in wild-type mice for maintaining LTP and memory. But another recent study (Wang et al. 2016) obtained conflicting results, with impaired maintenance of both late LTP and LTM following knockdown of PKMζ levels by a small hairpin RNA. Thus, there is contradictory evidence regarding the effects of inducible knockdown on maintenance of LTM. Another challenge for the PKMζ hypothesis of memory storage is the current lack of evidence for appropriate dynamic subcellular accumulation of PKMζ. Although PKMζ is observed to be strongly punctate in a subset of dendritic spines (Hernandez et al. 2013), no imaging studies have examined whether PKMζ accumulates specifically in dendritic spines (or presynaptic zones) that have been selectively potentiated. Thus, it would be important to develop a reporter that can dynamically monitor PKMζ levels or activity at potentiated synapses.

Overall, current evidence suggests a role for PKMζ in the late maintenance of at least some forms of memory. However, further work is necessary. In part, this necessity is to clarify the reasons for the differing results reported with inducible knockdowns. We also note a recent report (Rossato et al. 2019) that PKMζ is required for successful reconsolidation of an object recognition memory (ORM) following its reactivation, but is not required to maintain storage of an inactive ORM memory trace.

## Proposed maintenance of LTF, LTP, and LTM by persistent activation of the translation activator CPEB

A feedback loop based on persistent, direct upregulation of translation could maintain synaptic strength. For example, serotonin (5-HT)-induced LTF at sensorimotor synapses of *Aplysia* requires the translation activator CPEB (Si et al. 2003a), and it is suggested that local aggregation of CPEB can persistently increase translation of synaptic proteins at synapses that have undergone LTF (Li et al. 2018). LTF is similar to L-LTP in that LTF correlates with LTS, a form of LTM (Cleary et al. 1998; Frost et al. 1985). Inhibition of CPEB 24 h after LTF induction blocks the maintenance of LTF (Miniaci et al. 2008). CPEB has prion-like properties, forming stable synaptic aggregates (Si et al. 2003b). After transfection into yeast cells, CPEB aggregates self-perpetuate as distinct strains and aggregate a fusion protein with a CPEB domain (Heinrich and Lindquist 2011). Synaptic CPEB aggregation is increased by 5-HT, and an antibody that recognizes multimeric CPEB blocks LTF (Si et al. 2010). These results suggest that in *Aplysia*, CPEB aggregation may be self-perpetuating and necessary for maintaining LTF and LTM. Analogous results are found in *Drosophila*. Orb2, a CPEB homolog, forms prion-like aggregates, and their presence predicts memory strength (Li et al. 2016). Mutating Orb2 to inhibit aggregation (Majumdar et al. 2012) or inactivating Orb2 (Li et al. 2016) inhibits LTM. Aggregated Orb2 enhances translation (Khan et al. 2015).

This research has been extended to mammals. In mice, following neuronal stimulation, the CPEB3 isoform is activated and its aggregation increases (Drisaldi et al. 2015). CPEB3 promotes translation of GluA1 and GluA2 AMPAR subunits, which could enhance synaptic strength (Huang et al. 2006; Rayman and Kandel 2017). An inducible knockout of CPEB3 impairs maintenance of hippocampal LTP and two forms of spatial memory, and in addition, CPEB3 is not able to maintain LTP and LTM if its aggregation is prevented (Fioriti et al. 2015). After yeast transfection, CPEB3 aggregates self-perpetuate as distinct strains, similar to prions, and a specific CPEB3 domain is necessary for aggregate formation and for upregulation of translation of GluR2 mRNA (Stephan et al. 2015). Thus there self-perpetuating CPEB3 aggregation may play a role in maintaining LTP and LTM.

Studies with *Aplysia* and mammals have also addressed in part the possibility that CPEB aggregation in a prion-like manner might escape control, resulting in a spread of CPEB aggregates away from synaptic sites and throughout neurons. Specifically, CPEB aggregation and activity is regulated by several different processes. In *Aplysia*, a neuron-specific micro-RNA inhibits translation and thus aggregation of CPEB (Fiumara et al. 2015) and similar regulation occurs in mammals (Morgan et al. 2010). In mammals, CPEB3 aggregation is also regulated by SUMOylation (Drisaldi et al. 2015) and ubiquitination (Pavlopoulos et al. 2011).

However, some data at 3 d or more post-stimulus (Miniaci et al. 2008) argue against the hypothesis that CPEB aggregation and consequent local translation are important for maintenance of late LTF and LTM. Local synaptic application of either protein synthesis inhibitor, or CPEB antisense RNA, 72 h after induction of LTF, no longer blocked late maintenance of LTF assessed 120 h after induction. The authors suggested two interpretations of this failure to block: 1) After 72 h, CPEB-dependent local translation is no longer necessary to maintain LTF; or 2) CPEB-dependent local translation might still be necessary, but over even longer periods at a lower level. The second interpretation would imply that if CPEB activity was inhibited for a period on the order of days, and LTF then assayed, a significant impairment of LTF could occur. Further empirical investigation at late times, in *Aplysia* and in mammals, is therefore necessary to test the hypothesis.

Figure 2 summarizes synaptic positive feedback loops discussed in this review.

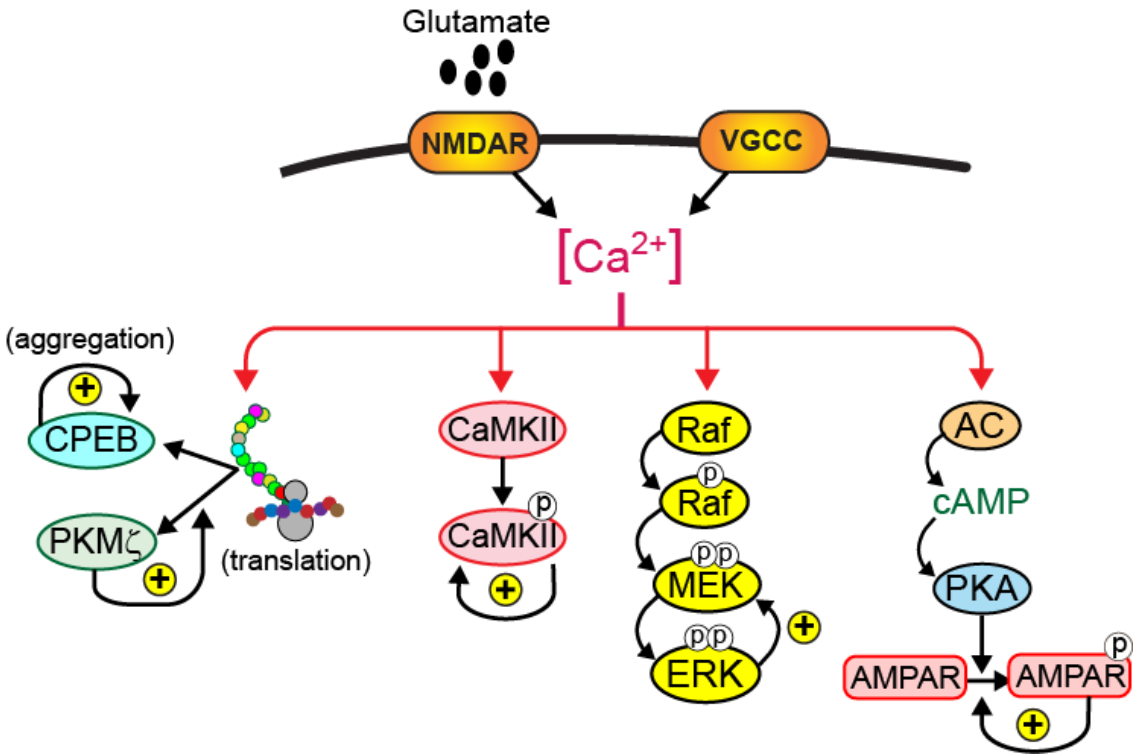

FIGURE 2. **Feedback loops based on self-sustaining kinase activation or translation, hypothesized to play roles in maintaining persistent LTP and LTM.** A stimulus (*e.g.*, tetani or theta-bursts) releases glutamate, leading to increased postsynaptic $Ca^{2+}$ *via* influx through NMDA receptors (NMDAR). Consequent depolarization leads to further $Ca^{2+}$ influx through voltage-gated $Ca^{2+}$ channels (VGCC). $Ca^{2+}$ binds to calmodulin (not shown) and activates, directly or indirectly, kinases and signaling pathways including CaMKII, the adenylyl cyclase – cAMP – PKA pathway, and the Raf – Mek – ERK pathway. Upstream processes in these pathways, such as Ras activation, are omitted for clarity. Local protein synthesis is also activated subsequent to $Ca^{2+}$ influx. Hypothesized positive feedback loops are indicated by plus signs. Within CaMKII holoenzymes, phosphorylated CaMKII subunits may phosphorylate and activate new or inactive subunits, persistently maintaining holoenzyme activity. The translational activator CPEB can aggregate in a self-perpetuating but regulated manner, possibly enabling a positive feedback loop that sustains increased translation. PKMζ is hypothesized to upregulate translation of its own mRNA, sustaining high PKMζ activity and forming another positive feedback loop. ERK has also been hypothesized to persistently upregulate its own activity, *via* phosphorylation and activation of its upstream activator, MEK kinase (Markevich et al. 2004; Smolen et al. 2008). Following stimulus-induced phosphorylation of AMPA receptors by PKA, AMPAR dephosphorylation may become saturated, allowing basal PKA activity to maintain elevated AMPAR phosphorylation (Hayer and Bhalla 2005). The activities of these kinases, and increased protein synthesis, may all cooperate to potentiate and maintain synaptic strength.

**Genetic and epigenetic positive feedback may be necessary to maintain long-term memory**

Empirical and modeling studies support the hypothesis that positive feedback in transcription may play an important role in maintaining memory. Two levels of feedback have been proposed. In one, transcription factors such as CREB upregulate their own transcription as well as that of other genes, resulting in enhanced protein synthesis, maintaining strong synapses and LTM. In the second level, epigenetic modifications, such as histone methylation/acetylation or DNA methylation, participate in self-sustaining feedback loops that upregulate transcription of genes that support memory. Although not discussed here, additional gene regulation, such as by non-coding RNAs, could also play important roles but have been less examined.

1. <u>Roles of positive feedback among transcription factors, and of growth factors, in maintaining LTF, LTP, and excitability</u>

In *Aplysia*, LTF is studied at the sensory neuron (SN) – motor neuron (MN) synapses that mediate withdrawal reflexes. Expression of CREB1 is necessary for LTF (Bartsch et al. 1998). In the SN, CREB1 is upregulated for at least 24 h after induction of LTF, and if this upregulation is blocked, LTF is inhibited (Liu et al. 2008, 2011). Activity of a related transcription repressor, CREB2, is concurrently downregulated in the SN, likely by ERK phosphorylation (Bartsch et al. 1995). These changes together upregulate genes that have cAMP response elements (CREs) in their promoters and are necessary for LTF. *Creb1* itself is such a gene, closing a positive feedback loop (Mohamed et al. 2005). *Creb2* also has CREs, thus cross-interaction between CREB1 and *creb2* forms a negative feedback loop in which CREB2 would, after upregulation, limit the induction of genes first induced by CREB1. Modeling has suggested these interlocked feedback loops could form a bistable switch such that a 5-HT application, which induces LTF, can switch *creb1* into a persistent state of high expression (Song et al. 2007). A caveat for generalizing this model to additional neuron types and training protocols is that the role of transcription factors may vary with cell type and with time post-stimulus. For example, at the sensorimotor synapse in a co-culture of an SN with an MN, persistent LTF lasting at least 7 days occurs after applying 5-HT on 2 consecutive days (Hu et al. 2015). In the MN and SN, CREB2 activity decreases 24 h after the second 5-HT application. However, in the MN, there is a late phase of CREB2 increase 48 h after 5-HT. Overexpression of CREB2 in the MN could replace the second 5-HT application in inducing persistent LTF. These data suggest that in the MN, at this later time, CREB2 may activate transcription to support

LTF. The analogous experiments were not done in the SN for CREB2, thus it is not known whether CREB2 might activate genes in the SN at this later time.

In *Drosophila*, overexpression of a CREB homologue, dCREB2a, was reported to activate transcription and to enhance LTM (Yin et al. 1995a). Subsequent experiments found a *dcreb2* mutation in this *Drosophila* line, suggesting that, apparently, normal dCREB2a did not enhance LTM (Perrazona et al. 2004). However, a later study found that induction of a 28 kDa isoform of dCREB2a does, in fact, enhance several forms of LTM (Tubon et al. 2013). dCREB2b is an alternatively spliced isoform of *dcreb2* and a transcription repressor (Yin et al. 1995b), and overexpression of dCREB2b inhibits LTM (Perrazona et al. 2004; Yin et al. 1994). Persistent induction, for several days, of a reporter gene driven by a CRE is observed in *Drosophila* following spaced training (Zhang et al. 2015). This persistent induction suggests a positive feedback loop may sustain activation of transcription driven by dCREB2a in *Drosophila*. CREB homologues also play roles in maintaining at least some forms of memory in the nematode *Caenorhabditis elegans* (Amano and Maruyama 2011) and in the honeybee (Gehring et al. 2016), although persistent CREB activation does not appear to have been examined in these species.

In mammals, CREB is necessary for at least some forms of LTM and L-LTP (Bourtchouladze et al. 1994; Bozon et al. 2003; Guzowski and McGaugh 1997; Josselyn et al. 2001; Kida et al. 2002; Pittenger et al. 2002; Sekeres et al. 2010). Strong synaptic input, inducing L-LTP, can transmit signals to the nucleus that upregulate gene expression and CREB activity. For example, the γ isoform of CaMKII translocates to the nucleus and activates CaM kinase IV, which phosphorylates CREB (Ma et al. 2014). Mutation of γCaMKII disrupts activity-dependent expression of genes necessary for L-LTP and impairs LTM (Cohen et al. 2018). Phosphorylated Jacob protein also translocates to the nucleus, leading to enhanced CREB activity (Karpova et al. 2013); and the CREB co-activator CRTC1 similarly translocates (Ch'ng et al. 2012). Somatic depolarization and action potential initiation, which can be elicited by strong synaptic input, also activates ERK, CREB, and immediate-early gene expression (Dudek and Fields 2002).

These data, however, mostly pertain to the induction of late LTP and LTM. Does a positive feedback loop also operate in mammals, maintaining CREB activation for many hours? Data suggests, but does not suffice to establish, that such feedback may play a role in later maintenance of at least some forms of memory. *Creb* has CREs in its promoter, which could facilitate persistent auto-activation of *creb* transcription (Meyer et al. 1993). Inhibitory avoidance (IA) learning leads to activation of hippocampal CREB persisting for at least 20 h, and transcription of the growth factor denoted brain-derived neurotrophic factor (BDNF) also persists at least 20 h (Bambah-Mukku et al. 2014). BDNF is required for IA learning (Bambah-Mukku et al. 2014) and LTP maintenance (Barco et al. 2005). CREB is required for BDNF transcription (Tao et al. 1998; Bambah-Mukku et al. 2014), and BDNF reciprocally mediates CREB activation (Pizzorusso et al. 2000). These reciprocal interactions constitute a positive feedback loop, although it is not yet known whether this loop is necessary for consolidation of LTM. Zhang et al. (2016) developed a model describing the ways in which this positive feedback loop may shape the dynamics of BDNF and CREB. The model simulates long lasting, although not bistable, CREB activation, and generates experimental predictions. These studies suggest that for IA learning and memory, positive feedback and consequent long-lasting CREB activation may be necessary. Further work is needed to generalize to other forms of mammalian learning, and to establish with a higher degree of confidence whether or not persistent activation of CREB or elevation of BDNF are important elements in maintaining diverse forms of late LTP and LTM. Mammalian studies do not currently provide specific evidence that a long-lasting increase in *creb* gene expression is necessary for consolidating or maintaining LTM.

ATF4, a mammalian homologue of *Aplysia* CREB2, is also important for the induction of late LTP and LTM. ATF4 commonly represses transcription (Karpinski et al. 1992). Decreased phosphorylation of eukaryotic translation initiation factor 2 $\alpha$ (eIF2$\alpha$) correlates with induction of hippocampal L-LTP, and in contrast, increased eIF2$\alpha$ phosphorylation has been correlated with increased expression of ATF4 (Costa-Mattioli et al. 2007). These data suggest relief of transcriptional repression, due to decreased eIF2$\alpha$ phosphorylation and ATF4 expression, may play a role in L-LTP. However, the role of ATF4 appears to be more complex. Knock down of ATF4 by RNA interference does not enhance, but rather impairs, hippocampal LTP and spatial memory (Pasini et al. 2015) and ATF4 can activate expression of the *parkin* gene (Bouman et al. 2011). Thus, further study is needed to delineate the roles of ATF4, including examination of the dynamics of ATF4 levels and activity following LTP induction, and possible effects of feedback interactions on those dynamics.

For *Aplysia* co-cultures of an SN with an MN, it is evidently reasonable that a persistent increase in gene expression, due to upregulation of CREB1, can sustain increased synaptic protein synthesis and synaptic weight for the single synapse. However, considering pyramidal neurons in mammalian cortex or hippocampus, which receive input from hundreds of other neurons, it appears much less likely that persistent upregulation of CREB activity and gene expression would correlate strongly with long-term maintenance of the strength of any one particular synapse. Tight synapse specificity would require that mRNAs are somehow marked, while in the soma, for delivery to one specific synapse out of a multitude. This mechanism has been denoted the "mail" hypothesis, and has been argued to be mechanistically inefficient (Frey and Morris 1998).

Nevertheless, neurons characterized by higher levels of CREB are preferentially incorporated into assemblies of neurons, commonly termed memory engrams, which are activated upon learning and upon memory retrieval (Park et al. 2016; Zhou et al. 2009). How can this be understood if upregulated CREB activity does not specifically enhance transport of gene products to synapses that store specific memories? One possible explanation relies on the observed correlation of increased CREB activity with increased neuronal excitability (Liu et al. 2011; Lopez de Armentia et al. 2007; Park et al. 2016; Zhou et al. 2009). Plausibly, more excitable neurons are more likely to display postsynaptic spiking and depolarization for a given stimulus, and thus a larger proportion of their afferent synapses would be enabled to undergo LTP (Mozzachiodi and Byrne 2010). In this case, neurons with higher CREB levels and activity would be more likely to be incorporated into engrams because their synapses have an increased, albeit nonspecific, likelihood of participating in LTP.

Feedback loops involving growth factors, sometimes in conjunction with CREB, may also play roles in maintaining synaptic strength, and in some cases induce gene expression. *Aplysia* LTF depends on a positive feedback loop in which ERK activation leads to activation of the TGF-β family of growth factors, which feed back to further activate ERK (Chin et al. 2006; Zhang et al. 1997). This ERK → TGF-β → ERK feedback is essential for a late phase of ERK activation in a two-trial training protocol, and consequent LTF (Kopec et al. 2015). Augmentation of LTF by ERK may be due in part to enhanced phosphorylation of CREB1 (Liu et al. 2017) and inactivation of CREB2 (Bartsch et al. 1995). In mammalian LTP, a similar feedback loop may be important. An inhibitor of TGF-β1 impairs LTP and object recognition memory (Caraci et al. 2015) and treatment of hippocampal neurons with TGF-β2 enhances CREB phosphorylation and evoked postsynaptic currents (Fukushima et al. 2007). Another positive feedback loop that may maintain *Aplysia* LTF involves the neurotrophin ApNT (Jin et al. 2018; Kassabov et al. 2013). PKA activation increases release of ApNT from SNs. ApNT binds to ApTrk receptors, and activation of PKA downstream of ApTrk closes a positive feedback loop (Jin et al. 2018). The resulting persistent PKA activation may help maintain CREB1 activity (Bartsch et al. 1998). Also necessary for LTF in *Aplysia* is the neuropeptide sensorin, the synthesis and secretion of which is regulated by PKA (Hu et al. 2006). Sensorin binds autoreceptors and activates MAPK post 5-HT treatment (Hu et al. 2004; Hu et al. 2006; Ormond et al. 2004; Sharma et al. 2003). Another *Aplysia* neurotrophin, ApCRNF, enhances neurite elongation and facilitates MAPK activation and LTF (Pu et al. 2014). More study is needed, however, to determine whether the sensorin and ApCRNF pathways form parts of closed positive feedback loops.

In mammals, nerve growth factor (NGF) is necessary for spatial learning and other forms of LTM (Conner et al. 2009). BDNF is necessary for persistence of inhibitory avoidance LTM assessed 7 d post-training (Bekinschtein et al. 2008). Several other growth factors regulate dendritic spine density, and dendritic length and complexity (Kopec and Carew 2013). Therefore, it is plausible that positive feedback loops involving growth factors and their regulation of intracellular signaling pathways play important roles in the maintenance of mammalian LTM. However, we note that for *Aplysia* and mammals, there is no current data concerning synapses or neurons involved in LTF or LTM, that indicates MAPK or PKA can maintain elevated activity over the time scales of days to months that characterize late memory maintenance.

**2.** Roles of persistent epigenetic modifications for maintaining LTF, LTP, and LTM

Studies have suggested epigenetic modifications are important for synaptic plasticity and memory. Histone acetylation is increased after fear conditioning in the hippocampus and amygdala, in part *via* recruitment of CREB together with its coactivator, CREB binding protein (CBP), which is a histone acetyltransferase (Vecsey et al. 2007). Correspondingly, inhibition of histone deacetylase enhances fear conditioning and LTP (Levenson et al. 2004; Monsey et al. 2011). Knockdown or mutation of CBP impairs LTP and LTM (Alarcon et al. 2004; Korzus et al. 2004; Wood et al. 2005) and impairs LTF (Liu et al. 2013). CBP mutation is responsible for Rubinstein-Taybi syndrome, a cause of intellectual disability (Petrij et al. 1995). Following fear conditioning, histone phosphorylation is also increased *via* the ERK pathway (Chwang et al. 2006). In *Aplysia* neurons, 5-HT application and consequent CREB activation induces histone acetylation that correlates with LTF (Guan et al. 2002).

DNA methylation is also upregulated in the hippocampus and amygdala after fear conditioning, and inhibition of DNA methylation blocks fear LTM (Miller and Sweatt 2007; Monsey et al. 2011). In *Aplysia*, DNA methyltransferase inhibition 24 h after LTS induction eliminates established LTS (Pearce et al. 2017). DNA methylation appears to also play a role in remote memory maintenance. A single associative learning trial induced hypermethylation in cortical neurons of rats that persisted for at least 30 days, and importantly, pharmacologic inhibition of methylation after 30 days disrupted LTM (Miller et al. 2010).

These data are intriguing in that DNA methylation correlates most commonly with repression of gene expression (Moore et al. 2013). How, then, could long-term methylation help maintain expression of plasticity-related proteins necessary for maintaining synaptic strength and LTM? Two, not mutually exclusive, possibilities are: 1) Less commonly, DNA methylation correlates paradoxically with activation of gene transcription (Chahrour et al. 2008; Kotini et al. 2011). Therefore, expression of some plasticity-related proteins may be directly enhanced by persistent hypermethylation. 2) DNA methylation represses expression of transcription factors that in turn repress expression of plasticity-related proteins, and DNA methylation may also repress expression of phosphatases that in turn repress activity of kinases important for maintaining LTM. Such dual repression following DNA methylation would yield net activation of molecular processes that may help maintain memory. Instances of dual repression have been reviewed (Kukushkin and Carew 2017). Fear conditioning increases DNA methylation and decreases expression of the Ser/Thr phosphatases PP1 and calcineurin for, respectively, at least 1 day and as long as 30 days after training (Miller and Sweatt 2007). This relief of phosphatase activity may enhance activities of kinases regulated by Ser/Thr phosphorylation, such as CaMKII and ERK. In *Aplysia*, the CREB2 promoter is methylated for at least 1 day after 5-HT application, diminishing *creb2* expression (Rajasethupathy et al. 2012). By relieving transcription repression due to CREB2, this methylation is likely to enhance and prolong expression of CREB1 or other genes suppressed by CREB2.

The above data suggest epigenetic mechanisms have substantial roles in maintaining several forms of LTM (Kim and Kaang 2017). For histone modifications, computational models have described mechanisms by which stable spatially restricted domains of modified nucleosomes could be sustained by positive feedback loops, persisting despite protein turnover (Dodd et al. 2007; Mukhopadhyay and Sengupta 2013). However, empirical studies supporting the maintenance of modifications by these feedback loops have to this point focused on yeast (Thon and Friis 1997; Obersriebnig et al. 2016), or other non-neuronal cells (Hathaway et al. 2012), rather than on neurons.

Although persistent epigenetic modification is an attractive mechanism for maintaining LTP and LTM, this nuclear mechanism encounters the same theoretical difficulty in ensuring synapse specificity as does regulation by transcription factors. But perhaps, as with transcription factors, epigenetic regulation acts largely through modifying neuronal excitability. DNA methylation does regulate excitability, and also regulates synaptic strength in a global manner. Inhibiting neuronal activity decreases DNA methylation and upscales the strength of glutamatergic synapses, and inhibiting DNA methylation similarly upscales glutamatergic synapses (Meadows et al. 2015), as well as increasing neuronal excitability (Meadows et al. 2016). Increased excitability or glutamatergic synaptic strength would yield greater postsynaptic depolarization for a given stimulus. Because of this greater depolarization, one plausible effect of upscaling synaptic strength (or neuronal excitability) would be to increase in a nonspecific manner the likelihood of LTP for afferent synapses, plausibly increasing the probability of incorporation into a memory engram.

Kyrke-Smith and Williams (2018) posit that persistent regulation of both transcription and the epigenome is necessary. These authors suggest maintenance of LTP and memory is reliant on persistent transcription of genes that act, in turn, to repress genes that promote synaptic plasticity. Thus, for neurons that are members of cell assemblies participating in memory storage, persistent transcription of repressor genes would inhibit further LTP or LTD, thereby tending to maintain these neurons in a state of relatively fixed synaptic strengths. Such a state would, by default, maintain previous LTP and memory. The authors suggest histone deacetylase 2 (*hdac2*) could be a persistent repressor gene, and discuss evidence that *hdac2* negatively regulates the ability of synapses to undergo structural plasticity. However, upregulation of *hdac2* appears to have only been examined up to 24 h after the induction of LTP (Ryan et al. 2012). In addition, for persistent upregulation of repressor genes to play a role in maintaining synaptic strength for weeks or months, this upregulation would need to be embedded within a positive feedback loop in order to maintain a stable state of elevated transcription. No such loop has been suggested. Therefore, while intriguing, the above hypothesis requires substantial further work to strengthen or falsify it.

In summary, upregulation of gene expression by transcription factors or epigenetic modification will lead to enhanced distribution of newly synthesized mRNAs and proteins throughout dendrites, where they would be available to maintain synaptic strength and memory. Figure 3 summarizes some of the mechanisms and proposed feedback loops that regulate gene expression and may be important for maintenance of LTM.

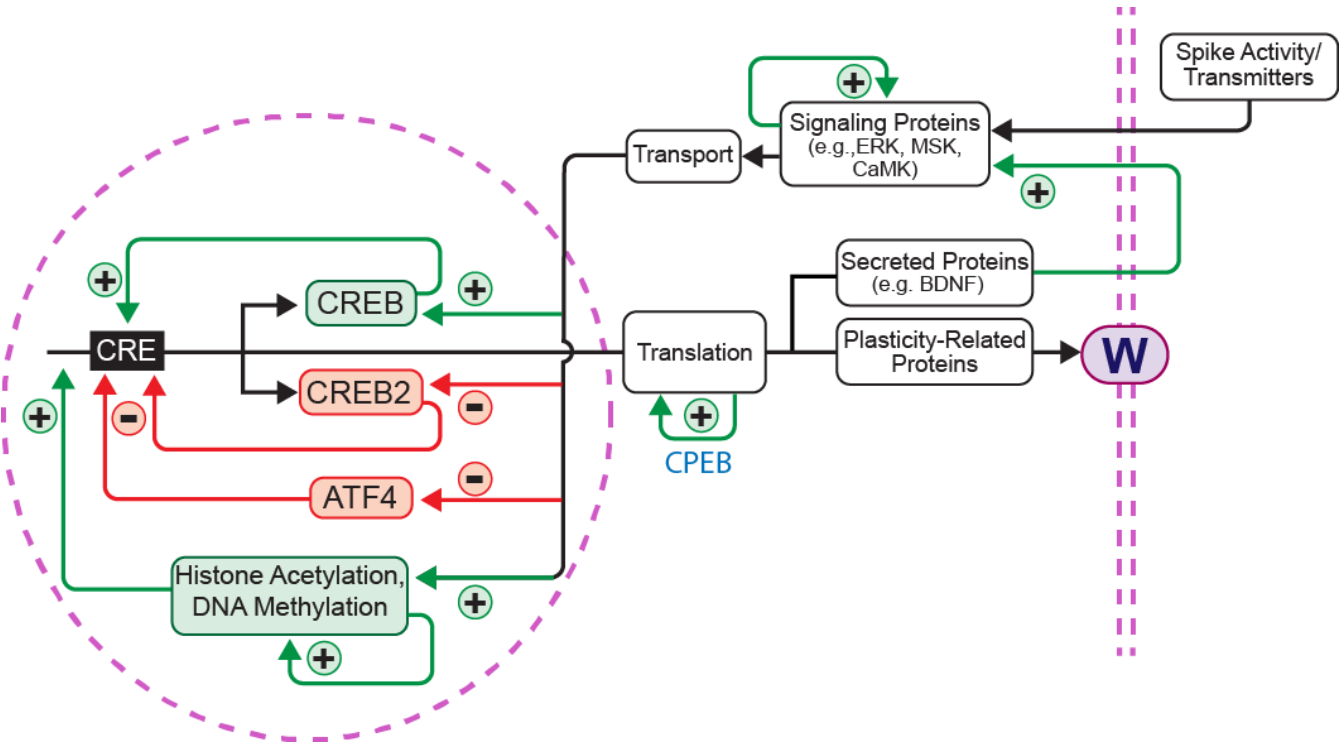

FIGURE 3. **Generic network of genetic and epigenetic regulatory mechanisms that may help maintain LTM**. Stimuli that induce late LTP signal from synapses to the nucleus. Signals may be transported by proteins such as Jacob (Melgarejo da Rosa et al. 2016), by activation of kinases such as ERK (Stough et al. 2015), or most directly by spread of action potentials, with consequent elevation of somatic and nuclear $Ca^{2+}$ levels and activation of signaling pathways (Dudek and Fields 2002). These pathways activate nuclear ERK and downstream kinases such as mitogen and stress-activated kinase (MSK) (Arthur et al. 2004), which activate CREB and other transcription factors and induce gene expression (shown as induction of a gene with a CRE in its promoter). Plus signs indicate putative positive feedback loops. Repression of gene expression, by transcription factors such as ATF4 in mammals or its homologue CREB2 in *Aplysia*, is also relieved, reinforcing expression. Minus signs indicate either repression or relief of repression. Although ATF4 is shown as a transcription repressor, recent work indicates it can also activate expression of some genes. Thus further work is needed to characterize the role of ATF4. The *creb* gene has a CRE, thus CREB can activate its own expression. Stimuli also activate histone-modifying enzymes such as CBP acetyltransferase, and DNA-modifying enzymes such as DNA methytransferases / demethylases. For histones in regulatory chromatin segments, models have suggested positive feedback may maintain domains of increased acetylation, facilitating persistent gene expression. Translation of mRNAs is regulated by factors such as CPEB. Upregulated translation of plasticity-related proteins (Frey and Morris 1998) is necessary to persistently strengthen synapses that participate in storage of LTM (increased synaptic weight W). Dashed purple circle denotes the nucleus, dashed double line denotes the cell membrane across which BDNF and other neurotrophins are secreted. These neurotrophins act through receptors (not shown) to activate kinases and signaling pathways.

**Preservation of LTM and late LTP appears to require ongoing synaptic reactivation**

An observation of key importance (Cui et al. 2004; Cui et al. 2005) suggests none of the positive feedback loops discussed above are individually sufficient to maintain long-term memories for a substantial part of a mammal's life. Inducible and reversible, forebrain-restricted, genetic knockdown of the NR1 subunit of the NMDA receptor severely disrupted retention of 8-9-month old contextual and cued fear memories (Cui et al. 2004). NR1 knockdown occurred 6 months after initial training. The NR1 knockdown blocks activity-induced $Ca^{2+}$ influx through the NMDA receptor, in turn blocking activation of downstream kinases (*e.g.*, CaMKII, ERK) that, taken together, are necessary to sustain synaptic translation as well as other processes (*e.g.*, AMPA receptor phosphorylation) that maintain strong synapses. After reversal of the knockdown, learning and memory in subsequent tasks was normal, illustrating a lack of permanent neuronal damage that would impair recall or performance. The NR1 knockdown also eliminated remote maintenance of nondeclarative taste memory (Cui et al. 2005) and eliminated retention of hippocampal LTM during the first weeks after training (Shimizu et al. 2000).

The above studies suggest periodic reactivation of assemblies of connected neurons, or engrams, that store specific memories is likely to be essential for preserving these memories, and presumably associated late LTP, over weeks or months. Periodic reactivation would induce $Ca^{2+}$ influx through the NMDA receptor, in turn reactivating the biochemical processes (kinase activation, local protein synthesis) that generated the original L-LTP. Thus, periodic reactivation would provide repeated "rounds" of LTP, reinforcing and thereby sustaining over months the original LTP, and the associated engram and memory. Synaptic reactivation is also likely, *via* signaling to the nucleus, to induce reactivation of CREB and other transcription factors. The resulting gene induction and global increase of synaptic protein levels may constitute an additional, permissive factor for reinforcement of LTP.

Additional data support the correlation of engram maintenance with late maintenance of LTP and with periodic neuronal reactivation. In contextual fear memory engrams, enhanced CA3-CA1 synaptic strength correlates with increased memory strength, and occludes LTP (Choi et al. 2018). Repeated spontaneous replay of engrams that encode recent experience is now well established (Ikegaya et al. 2004; Miller et al. 2014; Wu and Foster 2014). In particular, replay of recently learned engrams during sleep is frequent and is likely to contribute to engram consolidation (Giri et al. 2019; Wei et al. 2016). It was determined (Schapiro et al. 2018) that increased hippocampal replay of engrams encoding specific learned items correlated with improved memory of those items 12 h after learning. These studies describing replay suggest the synaptic activity occurring during engram reactivation resembles, in intensity or temporal properties, the activity that occurred during formation of the memory, supporting the hypothesis that reactivation may reinforce existing LTP of engram synapses. Properties of engrams have recently been further reviewed (Josselyn et al. 2017; Tonegawa et al. 2018).

Modeling studies are beginning to put this 'reactivation' hypothesis on a quantitative footing. A model hippocampal neural network able to store engrams (Wittenberg and Tsien 2002) incorporated positive feedback between synaptic reactivation and synaptic weight in the simplest manner, as a direct linear coupling between synaptic weight and the derivative of membrane potential. Although not bistable, this model sufficed to demonstrate that repeated reactivation, driven by spontaneous activity, could lead to selective consolidation and maintenance of a subset of engrams. A later model (Tetzlaff et al. 2013) combined synaptic homeostatic scaling (Turrigiano et al. 1998) with LTP and ongoing synaptic activity to provide a robust and stable model of memory maintenance based on repeated synaptic reactivation.

The experiments that support the necessity of synaptic reactivation for maintaining LTP and LTM do not yet demonstrate a complete positive feedback loop, because synapses that store LTP and memory have not been shown to be reactivated more frequently or strongly than other synapses, or to have LTP selectively and repeatedly reinforced. However, it appears likely that strengthened synapses that store memory would have greater numbers of NMDA receptors and thus, on average, exhibit larger elevations in $Ca^{2+}$ concentration due to ongoing spontaneous reactivation, leading to enhanced reactivation of intracellular signaling pathways that reinforce LTP. One modeling study (Smolen 2007) implemented positive feedback between synaptic weight and reactivation frequency in the context of a previous model describing stimulus-induced kinase activation and late LTP induction (Smolen et al. 2006). Bistability resulted, with late LTP preserved indefinitely. However, although this model represented intracellular signaling pathways, such as the ERK and PKA cascades, it was limited to describing dynamics of a few synapses converging onto a single neuron. It appears that no computational model has yet integrated neuronal network engram dynamics, synaptic reactivation, positive feedback, and biochemical events self-consistently. The complexity of such a model would be substantial.

Simulations with an attractor neural network model supported a variant of the reactivation hypothesis in which engrams are not explicitly reactivated (*i.e.*, the entire set of synaptic connections is not simultaneously reactivated) (Wei and Koulakov 2014). These authors argued that under these conditions, correlations in neural activity would still carry imprints of new and old engrams. Given appropriate parameter values for spike timing-dependent plasticity, these correlations could allow older engrams to be maintained by the network, by preferentially reinforcing the strength of synapses in those engrams. Within this "implicit rehearsal" mechanism for maintaining engrams, synaptic weight is larger for synapses with stronger correlations between pre- and postsynaptic activities. The authors suggest that the prediction of greater weight for synapses with correlated activity could be tested by simultaneous measurements of synaptic strength and ongoing activity. However, it may be in question how these experiments would clearly differentiate "implicit rehearsal" from a contrasting mechanism of "explicit rehearsal" in which entire engrams (or large portions) are reactivated, since explicit rehearsal should also correlate pre- and postsynaptic activities.

If ongoing synaptic reactivation maintains strong synapses and is also more frequent at strengthened synapses that are part of memory engrams, closing a positive feedback loop, empirical predictions follow. Selectively enhanced synaptic reactivation should lead to increased time-averaged activity of kinases implicated in LTP induction, in order to engage strengthening processes that counteract synaptic weight decay. These kinases include ERK, CaMKII, and plausibly PKMζ. Monitoring FRET-engineered substrates of these kinases over relatively long periods (hrs or longer) could demonstrate increased average activity. It also appears likely that the synaptic tag identified as being necessary for "capture" of plasticity-related proteins (PRPs), essential for late LTP (Frey and Morris 1997; Frey and Morris 1998) should be persistently and selectively set at strong synapses in order to allow periodic reinforcing capture of PRPs, *in vivo* or in active slice preparations. Sossin (2018) has discussed the likelihood that such a tag is maintained, noting that late tag maintenance, unlike initial tag setting, would probably depend in part on transcription. The molecular identity of the synaptic tag has yet to be established, but several species have been suggested to participate including the BDNF receptor TrkB (Lu et al. 2011), CaM kinase II (Redondo et al. 2010), and polymerized actin (Ramachandran and Frey 2009). One or more of these species is predicted to be persistently upregulated at strong synapses. Persistence of a molecular complex corresponding to a synaptic tag despite ongoing molecular turnover, selectively at engram synapses, could itself be argued to necessitate a positive feedback loop, in order to selectively maintain increased synthesis or trafficking of tag components.

It is established that recently activated memories are commonly subject to reconsolidation (Alberini and LeDoux 2013; Besnard et al. 2012; Lewis and Bregman 1973; Misanin et al. 1968; Morris et al. 2006; Nader and Hardt 2009; Sara 2000). That is, there is a period on the order of hours, following activation, during which memory is labile to degradation. Degradation occurs if molecular processes known to be important for some forms of LTP are disrupted. For example, reconsolidation of some forms of LTM is blocked by inhibiting protein synthesis, inhibiting CREB function, or inhibiting ERK or PKA (Tronson and Taylor 2007; Zhang et al. 2010). Results from different groups, however, indicate that these manipulations, applied for limited times, commonly fail to block reconsolidation. Block of reconsolidation is dependent on the type of LTM, the timing and extent of inhibition of molecular processes, and other empirical parameters (Tronson and Taylor 2007), as well as on the type of reactivating event, *e.g.* reinforced vs. non-reinforced (Alberini and LeDoux 2013), and on event duration (Pedreira and Maldonado 2003). Considering the hypothesis that periodic synaptic reactivation is essential for late maintenance of LTM, reactivation may induce periodic rounds of reconsolidation, accompanied by periodic lability of memory to degradation. Experiments might therefore find late, spontaneous and variable, intervals during which LTM can be degraded by temporarily disrupting molecular processes important for LTP. However, given the complex dependence of successful reconsolidation block on different empirical parameters and on the type of LTM, it does not appear possible yet to make firm empirical predictions about when and how established LTM would be disrupted in this manner, or to use such predictions to test the hypothesis that periodic synaptic reactivation is necessary for maintaining LTM. It appears that at this time, only a suggestion of an interesting direction for further research can be made.

**Crosstalk between feedback loops is likely to support rapid and stable memory storage**

We have discussed different positive feedback loops that may be necessary to maintain LTM, but do these loops operate independently, or are there data suggesting links, possibly reinforcing, between loops? For recurrent synaptic reactivation, there are such data. Synaptic activity stabilizes polymerized actin (F-actin), in part by recruiting the stabilizing factor profilin to dendritic spines (Basu and Lamprecht 2018). In this manner, synaptic reactivation would tend to stabilize enlarged spines against shrinkage. If stronger synapses do preferentially reactivate, then F-actin levels and synaptic reactivation would reciprocally reinforce each other in a positive feedback loop. BDNF promotes actin nucleation and remodeling (De Rubeis et al. 2013), and blocking BDNF signaling decreases spine F-actin levels and spine head size (Kellner et al. 2014). As discussed above, a second positive feedback loop may be formed by reciprocal activation of BDNF expression and persistent CREB phosphorylation (Bambah-Mukku et al. 2014).

Because BDNF appears to enhance spine F-actin and synaptic strength, this second loop could be linked with the F-actin – synaptic reactivation loop, supporting the hypothesis of reinforcing links between loops.

In addition, kinases hypothesized to participate in positive feedback loops affect the actin cytoskeleton. Activated CaMKII unbinds from F-actin, permitting access of actin regulatory proteins and actin remodeling and restabilization, important for structural plasticity of dendritic spines (Kim et al. 2015). ERK also enhances actin polymerization *via* a pathway that includes calpain activation (Zadran et al. 2010), suggesting possible crosstalk between feedback loops involving persistent CaMKII activation and ERK activation. Actin polymerization is also required for increasing translation of PKM$\zeta$, and would thus be essential for the hypothesized PKM$\zeta$ positive feedback loop (Kelly et al. 2007b). Also, as noted previously, PKM$\zeta$ increases trafficking of AMPARs to synapses. This process would act to increase excitatory postsynaptic currents and thus would be expected to increase the amplitudes of synaptic reactivation events. Finally, as discussed above, CaMKII phosphorylates CPEB, which in turn may upregulate translation of CaMKII, reciprocally linking proposed positive feedback loops dependent, respectively, on persistently elevated CaMKII activity and on aggregated, active CPEB. Figure 4 summarizes some of these modes of crosstalk between signaling pathways and feedback loops.

Overall, the above discussion suggests it is likely that several positive feedback loops act in concert, with each one a necessary component, for long-term maintenance of at least some forms of LTP and LTM. At least some of these loops are likely to share one or more common elements – in particular, synaptic reactivation, and regulation of actin cytoskeleton dynamics to promote dendritic spine remodeling.

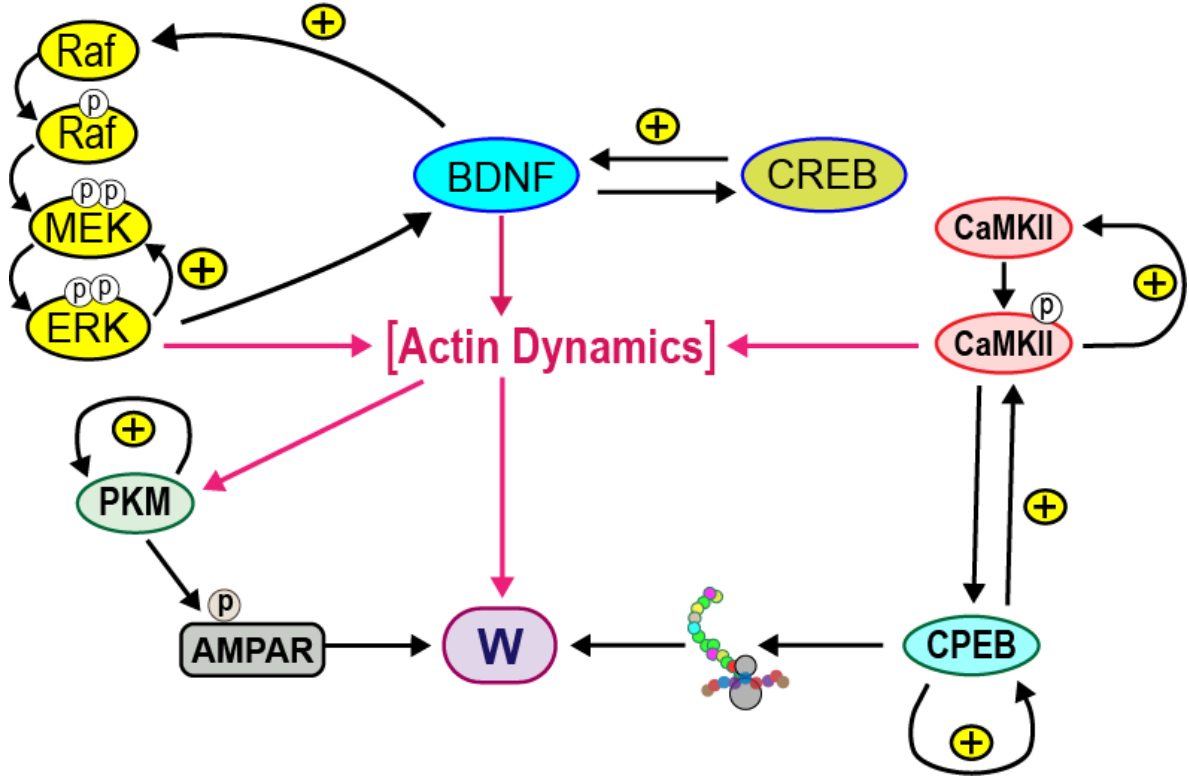

FIGURE 4. **Crosstalk between feedback loops hypothesized to maintain LTM.** Actin remodeling and polymerization (actin dynamics) is essential to increase and stabilize synaptic strength or weight (W). Actin dynamics are modulated by activation of ERK and CaMKII, as well as by increased BDNF. Activation of ERK, PKMζ, and CaMKII is hypothesized to be sustained by positive feedback loops (plus signs). BDNF expression is hypothesized to be activated by, and reciprocally enhance, CREB activation. BDNF expression is activated downstream of ERK (Zheng and Wang 2009), thus elevated BDNF expression could be sustained in part through sustained ERK activation. BDNF acts reciprocally, *via* the TrkB receptor, to activate the ERK pathway (Alonso et al. 2004), suggesting another positive feedback loop. Actin polymerization is required for increased translation of PKMζ, and is thus essential for hypothesized persistent PKMζ activation. The proposed synaptic feedback loop that sustains CaMKII autophosphorylation is interlocked with a putative feedback loop in which phosphorylated CPEB activates translation of α-CaMKII (Wu et al. 1998), with α-CaMKII reciprocally phosphorylating CPEB, maintaining elevated translation. Finally, self-perpetuating aggregation of CPEB is a putative positive feedback loop that may be interlocked with a CaMKII – CPEB feedback loop.

It is of further interest to consider studies that have used simplified computational models to examine the ways in which linked, reinforcing feedback loops may be advantageous for successful learning and memory. A two-loop model was simulated in which species A and B cooperate to enhance production of an output species OUT, which feeds back to enhance production of both A and B (Brandman et al. 2005). Here, OUT could represent synaptic strength and A and B could represent molecular species, such as kinases or translation factors, activated by synaptic activity. The time scale for changes of species A was set to be faster than that of species B. Two advantages of this dual-loop architecture were found. The faster feedback loop, between A and OUT, enabled a rapid response, with OUT rising quickly after stimulus. The slower loop between B and OUT increased the robustness of the response amplitude and shape. Its slow time constant filtered out stimulus fluctuations, decreasing fluctuations in OUT. This analysis was extended to a dual-loop model that exhibited bistability, so that OUT remained persistently elevated after stimuli (Zhang et al. 2007). The faster feedback loop drove a fast state transition, and the slower positive loop increased the stability of the basal and elevated states against stimulus fluctuations. It was further determined (Smolen et al. 2009) that such a dual-loop architecture increased the range of kinetic parameters allowing for bistability, and increased the stability of the basal and elevated states against stochastic fluctuations in molecule numbers. These studies suggest that multiple cooperating positive feedback loops give several advantages: a) rapid formation of memory, b) stability of LTM maintenance against fluctuations in molecule numbers due to protein/mRNA turnover, and c) robustness of LTM induction to variability in kinetic parameters and stimulus intensity.

**Alternative hypotheses for preservation of memory**

An alternative framework for preserving synaptic strength and LTM, not reliant on positive feedback, has been posited in which a cascade model of a synapse, with many hidden states, delinks memory lifetimes from signal response. This model enables quick learning in combination with slow forgetting (Fusi et al. 2005; Benna and Fusi 2016). The primary motivation for this model was to explain long-lasting stability of LTM despite continual storage of additional memories. In this framework, if a synapse is strong, and resides at a deep level of the hidden state cascade, its strength is metastable and decays very slowly (over months or longer). This interesting alternative remains to be tested empirically. At least two outstanding issues need to be addressed. First, what are proposed molecular or structural correlates of the currently abstract hidden synaptic states with long lifetimes? Second, can this framework account for the demonstrated requirement of ongoing synaptic reactivation and NMDA receptors to maintain remote memories? As another alternative, a model was proposed (Shouval 2005) in which clusters of interacting synaptic receptors, plausibly AMPA receptors, are enlarged by LTP induction, after which these clusters maintain their increased size for a long time in a metastable configuration, with a cluster lifetime far longer than the lifetime of individual receptor proteins.

**A network of interlocked feedback loops, at multiple levels, is likely to contribute to preserving LTM**

A complex network of positive feedback loops, ranging in scale from auto-activation of a single kinase, through activation of gene expression, and up to ongoing reactivation of networks or engrams of neurons, is likely to contribute collectively to the long-term maintenance of LTM. Figure 5 constitutes one representation of elements of this network and their interactions.

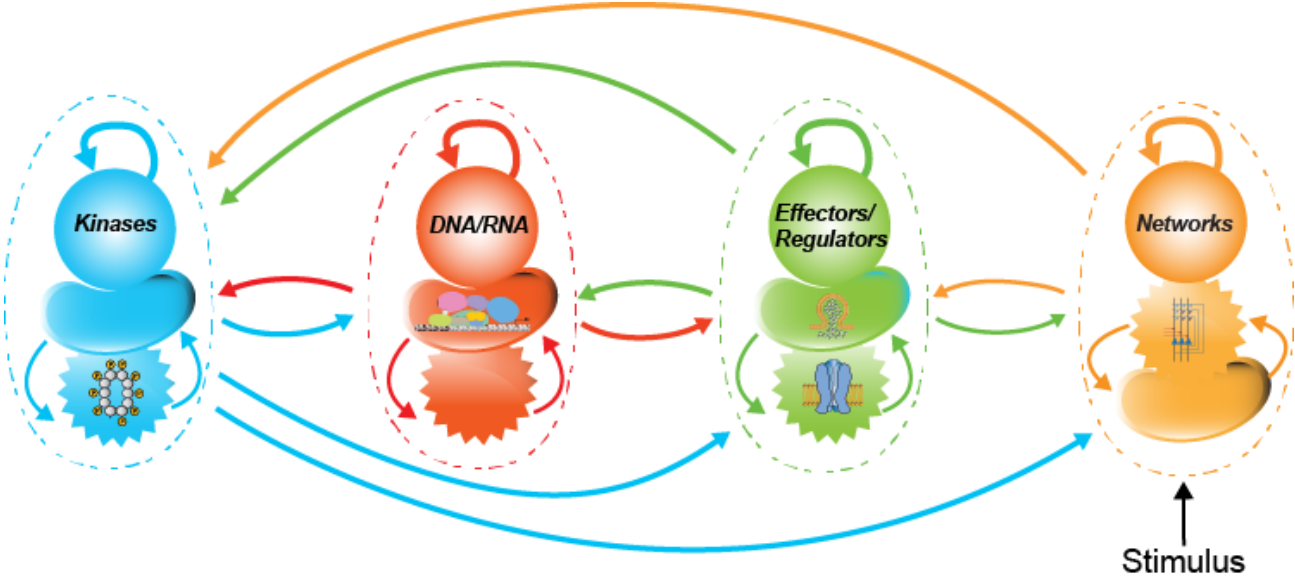

FIGURE 5. **Organization of classes of feedback loops hypothesized to maintain LTM**. A stimulus activates neuronal circuits (bottom right), with subsequent spike activity and transmitter release activating kinases at the level of synapses (left column). Some kinases may remain persistently active, either by autoactivation (CaMKII) or positive feedback between kinases (*e.g.*, ERK ↔ Raf, Balan et al. 2006). At the level of a neuron, synaptic activity induces gene expression, which may be prolonged due to autoactivation of transcription factor genes (in particular *creb*), or due to prolonged activation of kinases such as ERK that phosphorylate transcription factors (*e.g.*, Elk, CREB), inducing transcription. In each vertical column, the varying shapes point to the diversity of elements within a level (*e.g.*, diverse genes and kinases). At the level of secretion from a neuron and effects on itself or neighboring cells, neurotrophins such as BDNF, TGF-β, and ApNT act through receptors such as TrkB or ApTrk, activating kinases including ERK and phosphoinositide 3-kinase (Yoshii and Constantine-Paton 2010; Zadran et al. 2010). In turn, kinase activity can enhance neurotrophin release (*e.g.*, ERK → BDNF) generating positive feedback. Finally, at the level of neuronal networks, repeated reactivation of engrams appears to play a key role in memory maintenance. For example, in the hippocampal CA3 region, recurrent neural circuitry appears designed to facilitate re-entrant circuit and synaptic reactivation, enabling positive feedback and plausibly reinforcing synaptic strength and LTM (Rebola et al. 2017). More generally, recurrent neural circuit architecture is common in the neocortex (Douglas and Martin 2007).

As was pointed out early on (Schwartz and Greenberg 1989), the complexity of these interactions, still far from being fully understood, reminds one of an Indian story about how celestial bodies are organized, as a never-ending series of "turtles all the way down" (Geertz 1973). Given the plethora of interacting feedback loops suggested by recent data, an update might be "turtles all the way down and back again". Evidently, there is rich ground here for experiments and modeling to delineate these interactions and their dynamic consequences for formation, modulation, and preservation of LTM.

Finally, it is of particular interest to consider which subset of data helps to delineate, specifically, mechanisms that are likely to preserve LTM on the remote time scale of weeks, months, or years. For some of the putative feedback loops discussed above, such data are lacking. For example, long-lasting activation of ERK is an essential component of some proposed feedback loops (Bhalla and Iyengar 1999; Smolen et al. 2008) including putative feedback loops involving growth factors. However, persistent ERK activation, lasting weeks or more, has not been observed at specific synapses or engram neurons. Similarly, although data have shown that neurons overexpressing CREB are preferentially recruited into engrams (Sehgal et al. 2018), persistently enhanced endogenous activity of CREB, on time scales of weeks, has not been demonstrated in engram neurons. Intriguingly, it appears that on these time scales, current data focus on three feedback loops we have discussed: 1) increased PKMζ activity, 2) persistent epigenetic modification, and 3) ongoing synaptic reactivation. In particular, the findings that: 1) PKMζ levels are increased for at least a month after spatial learning (Hsieh et al. 2017), 2) inhibition of DNA methylation 30 days after training inhibits LTM retention (Miller et al. 2010) and 3) synaptic reactivation appears necessary to maintain memory at later times, ~ 6 months after learning (Cui et al. 2004; Cui et al. 2005) support the importance of these proposed feedback mechanisms for preserving remote LTM.

## Acknowledgements

Supported by NIH grants NS019895 and NS102490. We thank C. Alberini, T. Carew, N. Kukushkin, H. Shouval, and Y. Zhang for comments on an earlier version of the manuscript.

## Declaration of Interests

The authors declare no competing interests.